\documentclass{PoS}
\usepackage{amssymb}
\usepackage{amsmath}
\title{Quantification of Chiral Magnetic Effect from Event-by-Event Anomalous-Viscous Fluid Mechanics}

\ShortTitle{Quantification of CME from Event-by-Event Anomalous-Viscous Fluid Mechanics}

\author{\speaker{Shuzhe Shi}\\
       Physics Department and Center for Exploration of Energy and Matter,
Indiana University, 2401 N Milo B. Sampson Lane, Bloomington, IN 47408, USA.\\
       E-mail: \email{shishuz@indiana.edu}}

\author{Yin Jiang\\
        School of Physics and Nuclear Energy Engineering, Beihang University, Beijing 100191, China.\\
        Institut f\"ur Theoretische Physik, Universit\"at Heidelberg, Philosophenweg 16, D-69120 Heidelberg, Germany.}

\author{Elias Lilleskov\\
        Department of Physics and Astronomy, Macalester College, 
1600 Grand Avenue, Saint Paul, MN 55105, USA.}

\author{Jinfeng Liao\\
Physics Department and Center for Exploration of Energy and Matter,
Indiana University, 2401 N Milo B. Sampson Lane, Bloomington, IN 47408, USA.\\
Institute of Particle Physics and Key Laboratory of Quark \& Lepton Physics (MOE), Central China Normal University, Wuhan, 430079, China.}

\abstract{Chiral Magnetic Effect (CME) is the macroscopic manifestation of the fundamental chiral anomaly in a many-body system of chiral fermions, and emerges as anomalous transport current in hydrodynamic framework. Experimental observation of CME is of great interest and significant efforts have been made to look for its signals in heavy ion collisions. Encouraging evidence of CME-induced charge separation has been reported from both RHIC and LHC, albeit with ambiguity due to potential background contributions. Crucial for addressing such issue, is the need of quantitative predictions for both CME signal and the non-CME background consistently, with sophisticated modeling tool. In this contribution we report a recently developed Anomalous Viscous Fluid Dynamics (AVFD) framework, which simulates the evolution of fermion currents in QGP on top of the data-validated VISHNU bulk hydro evolution. In particular, this framework has been extended to event-by-event simulations with proper implementation of known flow-driven background contributions. We report quantitative results from such simulations and evaluate the implications for interpretations of current experimental measurements. Finally we give our prediction for the CME signal in upcoming isobaric collisions.}

\FullConference{Critical Point and Onset of Deconfinement\\
                 7-11 August, 2017\\
                 The Wang Center, Stony Brook University, Stony Brook, NY}

\begin{document}

\section{Introduction -- Chiral Magnetic Effect \& Anomalous-Viscous Fluid Dynamics}
\label{intro}
The Chiral Magnetic Effect (CME)~\cite{Kharzeev:2004ey,Kharzeev:2015znc} refers to the generation of an electric current    $\vec{J}_Q$ along the {\it magnetic field} $\vec{B}$ applied to a system of chiral fermions with chirality imbalance, i.e.
\begin{eqnarray} \label{eq_cme}
\vec{ J}_Q = \sigma_5 \vec{ B}
\end{eqnarray}
where $\sigma_5 = C_A \mu_5$ is the chiral magnetic conductivity, with the chiral chemical potential $\mu_5$ that quantifies the imbalance between fermion densities of opposite (right-handed, RH versus left-handed, LH) chirality. 

Given the magnificent physics embodied in the Chiral Magnetic Effect, it is of the utmost interest to search for its manifestation in the quark-gluon plasma (QGP) created in relativistic heavy ion collisions at the Relativistic Heavy Ion Collider (RHIC) and the Large Hadron Collider (LHC). Dedicated searches for potential CME signals have been ongoing at RHIC and the LHC~\cite{STAR_LPV1,STAR_LPV_BES,ALICE_LPV,Khachatryan:2016got}, with encouraging evidences reported through measuring the charge separation signal induced by the CME current (\ref{eq_cme}). The interpretation of these data however suffers from backgrounds arising from the complicated environment in a heavy ion collision (see e.g.~\cite{Kharzeev:2015znc,Liao:2014ava,Bzdak:2012ia}). Currently the most pressing challenge for the search of CME in heavy ion collisions is to clearly separate background contributions from the desired signal. A mandatory and critically needed step, is to develop state-of-the-art modeling tools that can quantify CME contribution in a realistic heavy ion collision environment. 

To address this challenge, we've recently developed a simulation framework, the Anomalous-Viscous Fluid Dynamics (AVFD) \cite{Jiang:2016wve, Shi:2017cpu}, focusing on describing anomalous chiral transport in heavy ion collisions at  high beam energy (such as the top energy RHIC collisions). The bulk evolution in such collisions is well described by boost-invariant 2+1D 2nd-order viscous hydrodynamics (e.g. VISHNU simulations~\cite{Shen:2014vra}) where net charge densities are small enough and typically neglected without much influence on bulk evolution. However to study the CME, one needs to accurately account for the evolution of fermion currents. Our approach is to solve the following fluid dynamical equations for the chiral fermion currents (RH and LH currents for u and d flavors respectively) as perturbations on top of the bulk fluid evolution: 
\begin{eqnarray} 
\begin{aligned}[c]
\hat{D}_\mu J_{\chi, f}^\mu &= \chi \frac{N_c Q_f^2}{4\pi^2} E_\mu B^\mu \\
J_{\chi, f}^\mu &= n_{\chi, f}\, u^\mu + \nu_{\chi, f}^\mu + \chi \frac{N_c Q_f}{4\pi^2} \mu_{\chi, f} B^\mu  \\ 
\Delta^{\mu}_{\,\, \nu} \hat{d} \left(\nu_{\chi, f}^\nu \right) &= - \frac{1}{\tau_{r}} \left[  \left( \nu_{\chi, f}^\mu \right) -  \left(\nu_{\chi, f}^\mu \right)_{NS} \right ] \\
\left(\nu_{\chi, f}^\mu \right)_{NS} &=  \frac{\sigma}{2} T \Delta^{\mu\nu}   \partial_\nu \left(\frac{\mu_{\chi, f}}{T}\right) +  \frac{\sigma}{2} Q_f E^\mu   \quad
\end{aligned}
 \label{eq_avfd}
\end{eqnarray} 
where $\chi=\pm1$ labels chirality for RH/LH currents and $f=u,d$ labels light quark flavor with  electric charge $Q_f$ and color factor $N_c=3$. The $E^\mu=F^{\mu\nu}u_\nu$ and  $B^\mu=\frac{1}{2}\epsilon^{\mu\nu\alpha\beta}u_\nu F_{\alpha\beta}$ are   external electromagnetic fields in fluid rest frame.  The derivative $\hat{D}_\mu$ is covariant derivative and  $\hat{d} =u^\mu \hat{D}_\mu$, with projection operator $\Delta^{\mu \nu}=\left(g^{\mu\nu} - u^\mu u^\nu \right)$.  Viscous parameters $\sigma$ and $\tau_r$ are diffusion constant and relaxation time respectively. 
In the above equations the fluid four-velocity field $u^\mu$, temperature $T$ as well as all other thermodynamic quantities are determined by background bulk flow. Furthermore the (small) fermion densities $n_{\chi, f}$ and corresponding chemical potential $ \mu_{\chi, f}$ are related by lattice-computed quark number susceptibilities $c_2^f(T)$.

\section{Results from Smooth AVFD simulation}

As the key ingredients for driving the CME current, the magnetic field and the initial axial charge density are the most important inputs for the AVFD simulation. 
For the magnetic field $\vec {B} = B(\tau) \hat{y}$ (with $\hat{y}$ the event-wise out-of-plane direction), we use a plausible parametrization (see e.g.~\cite{Yin:2015fca}) as   
$B(\tau) = \frac{B_0}{1+\left(\tau / \tau_B\right)^2}$.
The peak value $B_0$ (for each centrality) at the collision point has been well quantified with event-by-event simulations and we use the most realistic values from \cite{Bloczynski:2012en}. For the lifetime of the $B$ field we use a reasonable estimate of $\tau_B=0.6$ fm/c which is  comparable to the onset time of hydrodynamic evolution. 
For the initial axial charge density arising from gluonic topological charge fluctuations, one could make the following estimate based on the strong chromo-electromagnetic fields in the early-stage glasma similarly to the recent study in \cite{Hirono:2014oda}: 
$ \sqrt{ \left< n_5^2 \right> } \simeq  \frac{Q_s^4\, (\pi \rho_{tube}^2 \tau_0) \, \sqrt{N_{coll.}}}{16\pi^2 \, A_{overlap}} $.
In the above $\rho_{tube}\simeq 1 \rm fm$ is the transverse extension of glasma flux tube, $A_{overlap}$ is the geometric overlapping area of the two colliding nuclei, and $N_{coll.}$ the binary collision number for a given centrality. Such axial charge density depends most sensitively upon the saturation scale $Q_s$, with a reasonable range of  $Q_s^2\simeq 1\sim 1.5 \rm GeV^2$ for RHIC 200AGeV collisions. 

\begin{figure}[!hbt]
\includegraphics[width=0.4\textwidth]{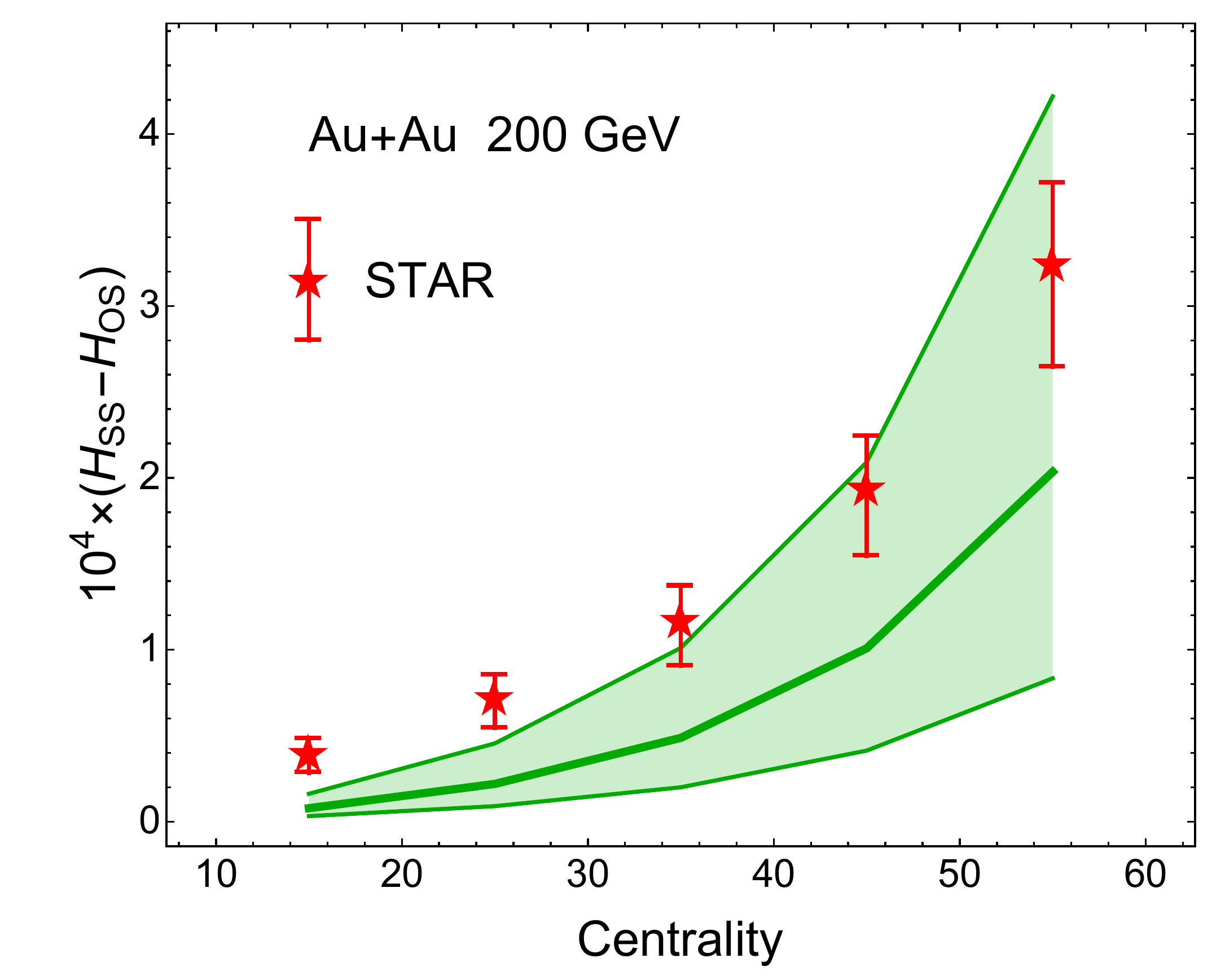} \hspace{1pc}%
\begin{minipage}[b]{20pc}
\begin{center} 
\caption{(color online)\label{fig.smooth}
 Quantitative predictions from Anomalous-Viscous Fluid Dynamics simulations for the CME-induced H-correlations, in comparison with STAR measurements~\cite{STAR_LPV_BES}. The uncertainty of experimental data comes from the uncertainty of $\kappa$. Central values correspond to $\kappa=1.2$, while upper and lower bounds are for $\kappa=1$ and $\kappa=1.5$, respectively. The green bands reflect current theoretical uncertainty in the initial axial charge generated by gluonic field fluctuations.}
\end{center}
\end{minipage}
\end{figure}

After the preceding discussions on the various aspects of the AVFD tool, let us now proceed to utilize this tool for quantifying CME signal to be compared with available data. The measurement of a CME-induced charge separation is however tricky, as this dipole flips its sign  from event to event depending the sign of the initial axial charge arising from fluctuations, thus with a vanishing event-averaged mean value. What can be measured is its variance, through azimuthal correlations for same-sign (SS) and opposite-sign (OS) pairs of charged hadrons. The so-called $\gamma_{SS/OS}\equiv \left< \cos(\phi_1+\phi_2)\right>$ observables measure a difference between the in-plane versus out-of-plane correlations and are indeed sensitive to potential CME contributions. They however suffer from considerable flow-driven background contributions that are not related to CME (see e.g. \cite{Kharzeev:2015znc,Bzdak:2012ia}). One plausible approach to  separate background and CME signal is based on a two-component scenario~\cite{Bzdak:2012ia}, which  was recently adopted by the STAR Collaboration to suppress backgrounds and   extract the flow-independent part (referred to as $H_{SS/OS}$)~\cite{STAR_LPV_BES}. We consider $H_{SS/OS}$ as our ``best guess'' thus far for potential CME signal to be compared with AVFD computations. Specifically a pure CME-induced charge separation will contribute as $\left(H_{SS}-H_{OS}\right) \to 2\left(a^{ch}_1\right)^2$. The AVFD results for various centrality bins are presented in Fig.~\ref{fig.smooth}, with the green band spanning the range of key parameter $Q_s^2$ in $1\sim 1.5 \rm GeV^2$  reflecting uncertainty in estimating the initial axial charge. Clearly the CME-induced correlation is very sensitive to the amount of initial axial charge density as controlled by $Q_s^2$. The comparison with STAR data~\cite{STAR_LPV_BES} shows very good agreement for the magnitude and centrality trend for choices with relatively large values of $Q_s^2$.

\section{Event-by-Event AVFD simulations}

\begin{figure}[!hbt]\centering
\includegraphics[width=0.4\textwidth]{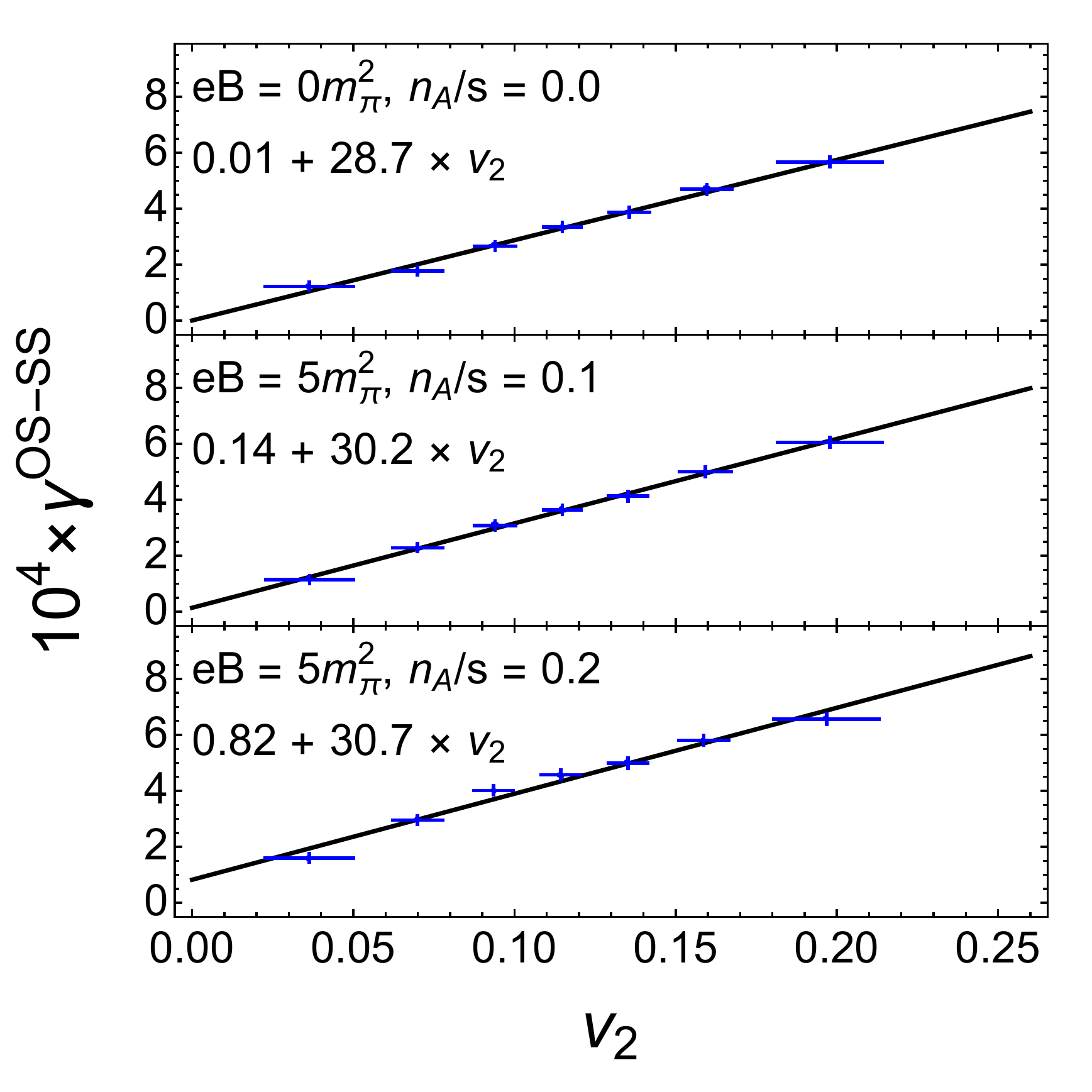}\quad
\includegraphics[width=0.4\textwidth]{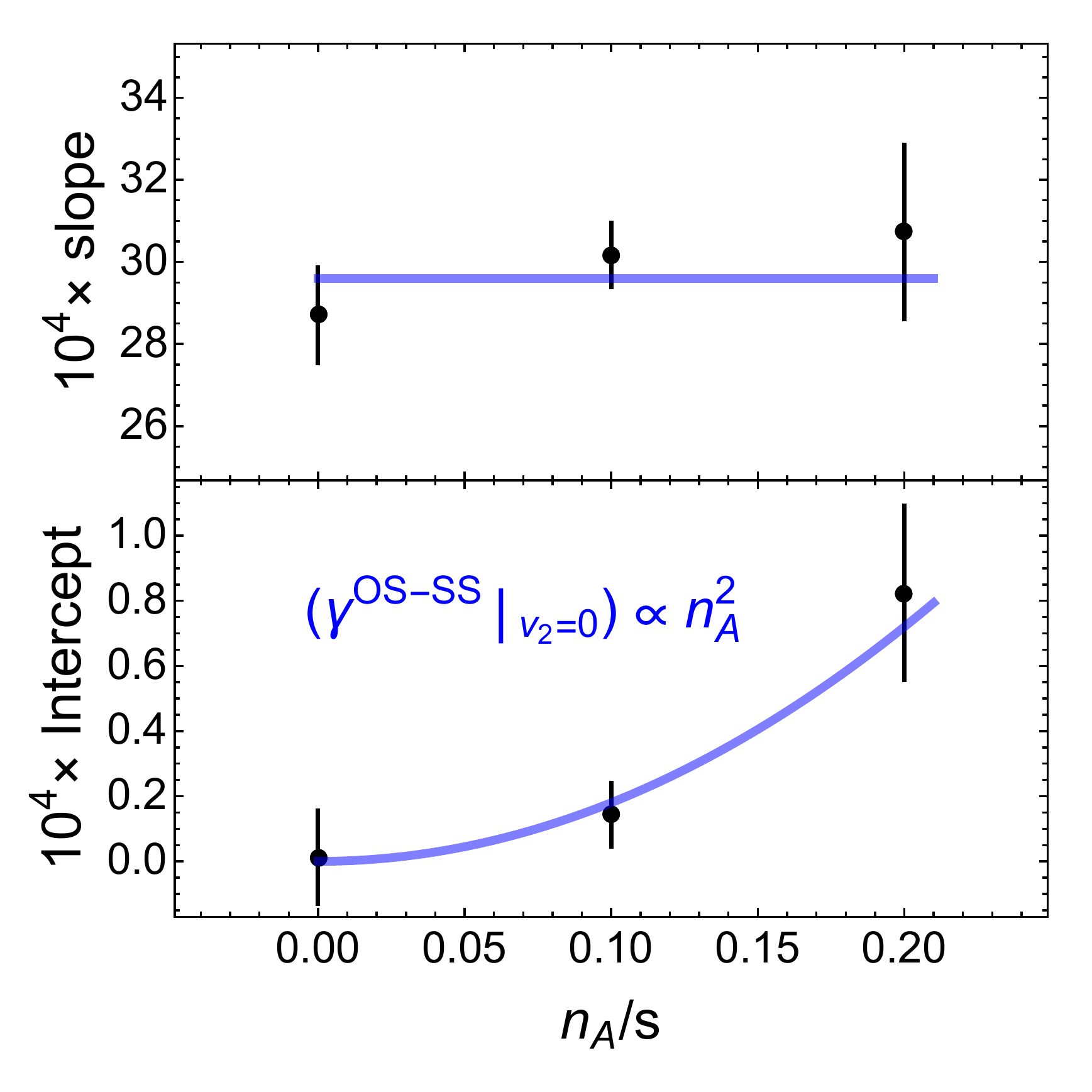}
\caption{(color online) (left) The $v_2$ dependence of $\gamma^{OS-SS}$; (right)  The slope and intercept extracted from the $v_2$ dependence of $\gamma^{OS-SS}$.}
\label{fig.EbE}
\end{figure}

While the AVFD simulations based on smooth average hydro profile provide a quantitative account of the ``pure'' CME signal, it is imperative to perform event-by-event simulations that could allow direct calculation of two-particle correlations and thus ultimate comparison with experimental measurements. We've recently implemented such simulations with event-wise fluctuating initial conditions as well as a hadronic stage after hadronization using UrQMD  to include the effects of hadron cascade and resonance decay. To demonstrate the CME-driven correlations, we  run simulations for $50-60\%$ 200GeV Au-Au collisions with three different setups: (a) a ``null'' case with no magnetic field or chiral imbalance; (b) $eB=5m_\pi^2$ and $n_A/s=0.1$ (corresponding to estimates with $Q_s^2\simeq 1 \rm GeV^2$); (c) $eB=5m_\pi^2$ and $n_A/s=0.2$ (corresponding to estimates with $Q_s^2\simeq 1.4 \rm GeV^2$). For each setup, we accumulated $\sim10^{7}$ events and measured the correlator $\gamma^{OS-SS}$ versus different event-wise $v_2$ bins, shown in Fig.\ref{fig.EbE} (left). An obvious linear dependence $\gamma^{OS-SS}$ on $v_2$ is observed for all these three cases. The extracted slope and intercept of such linear dependence are shown in Fig.\ref{fig.EbE} (right): while  the slope is basically independent of initial axial charge (indicating its dominant non-CME origin), the intercept grows quadratically with initial axial charge as expected from the CME signal.

\section{AVFD predictions for upcoming IsoBar Program}

Finally we report the AVFD predictions for the isobaric collisions planned at RHIC.
As has been intensively emphasized, the major difficulty of confirming the CME signal is our lacking of knowledge of the background, and unable to tell the CME signal from the non-CME background. To avoid such ambiguity, the IsoBar Program at RHIC\cite{Skokov:2016yrj} has been proposed to disentangle the CME signal, which collides $^{96}_{44}$Ru-$^{96}_{44}$Ru versus $^{96}_{40}$Zr-$^{96}_{40}$Zr systems.
Both of these nuclei have the 96 nucleons but their electric charge differs for $10\%$. The former means the similar bulk background in such system, while the latter means difference in magnetic field, which leads to different CME signal. By measuring the difference of $\gamma$- / $\delta$-correlators between these two systems, one can unambiguously decipher CME from backgrounds.

\begin{figure}[!hbt]\centering
\includegraphics[width=0.3\textwidth]{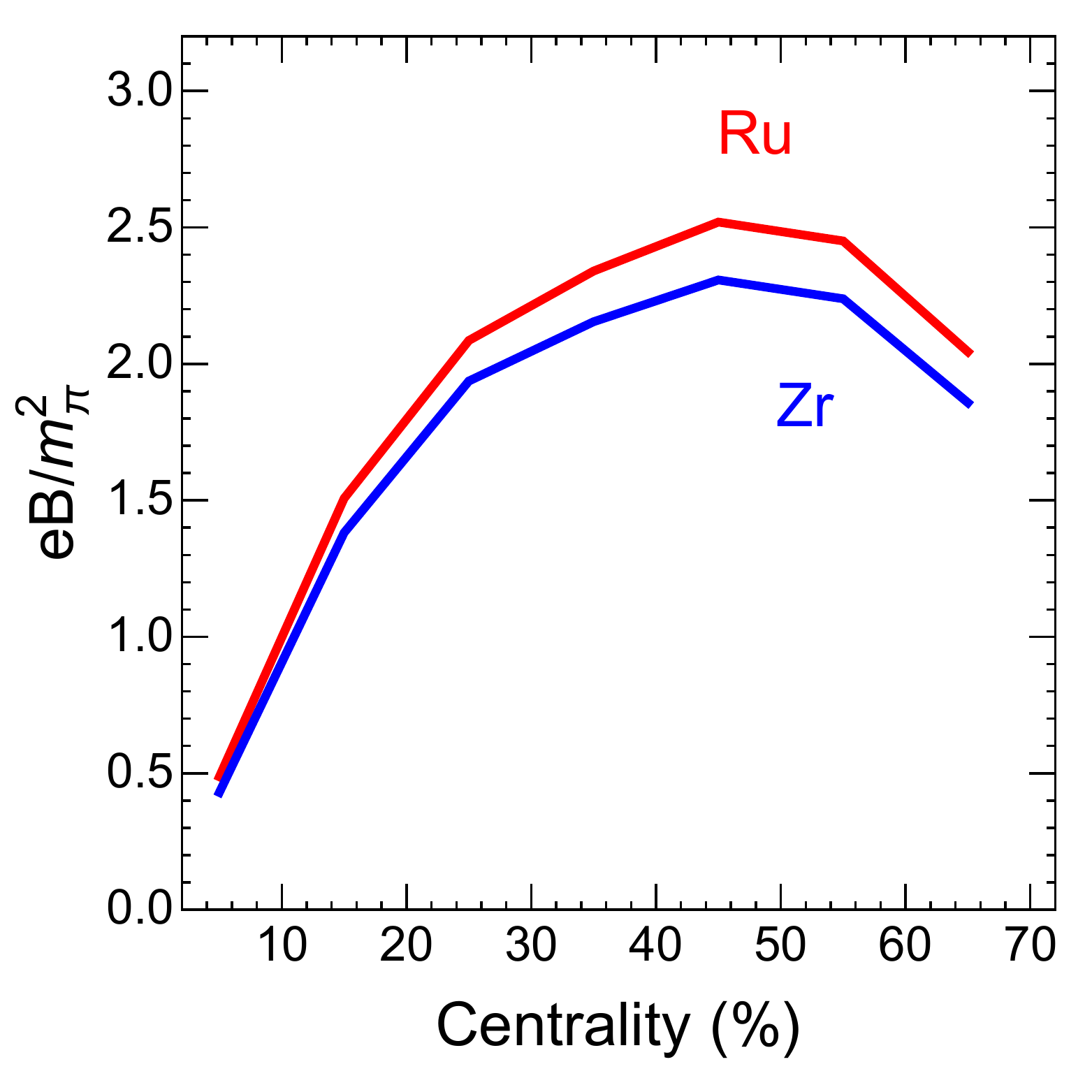}\quad
\includegraphics[width=0.3\textwidth]{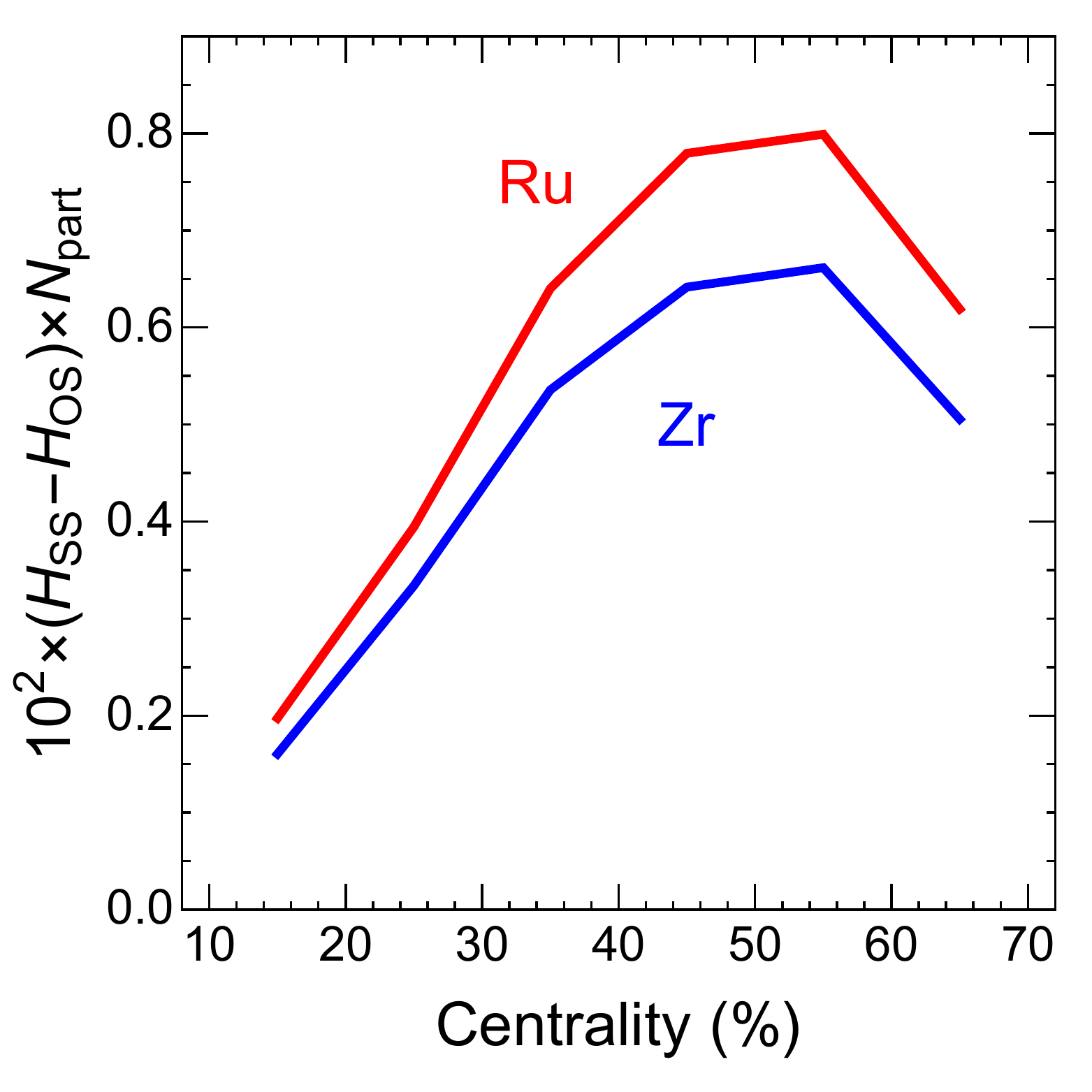}\quad
\includegraphics[width=0.3\textwidth]{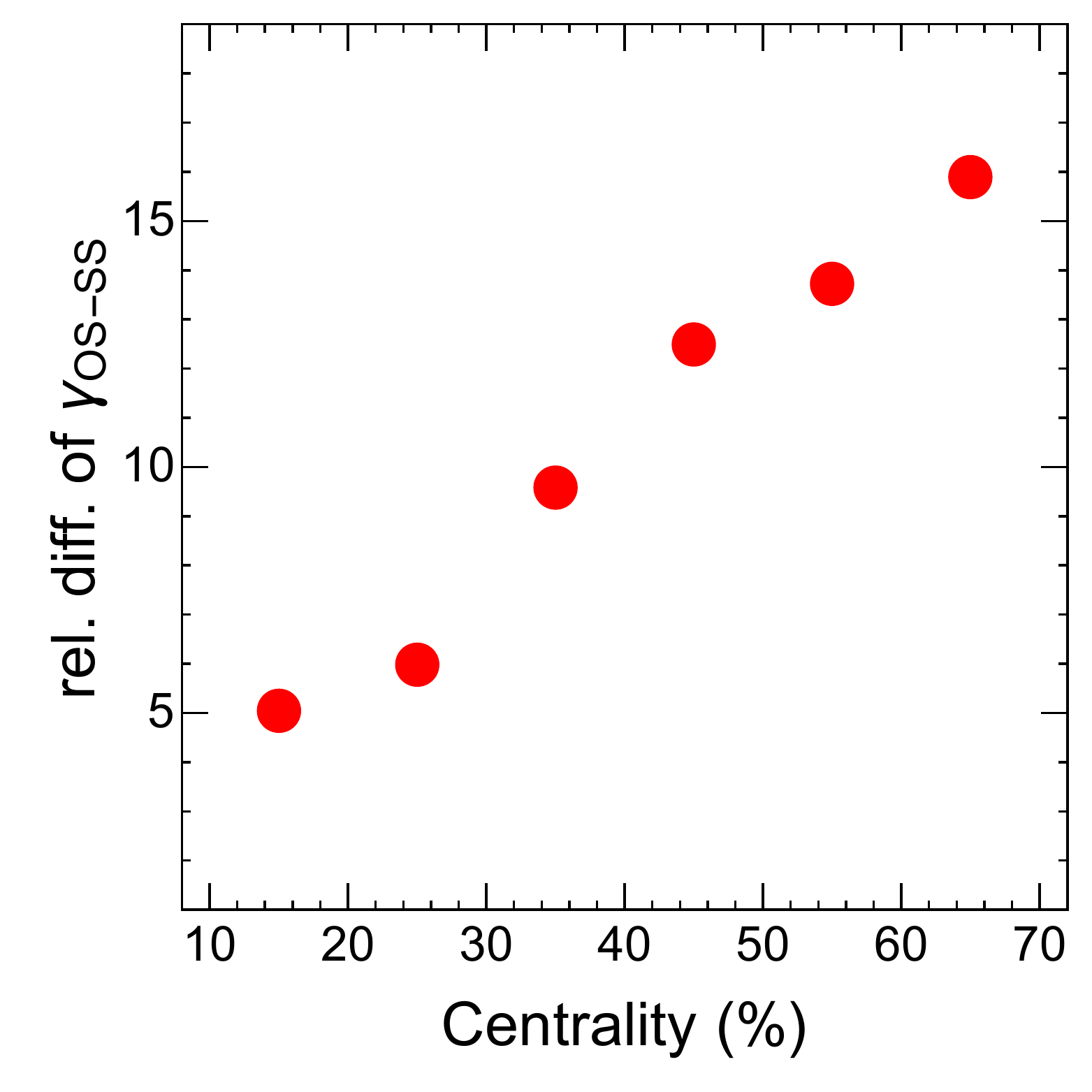}
\caption{(color online) (left) Projected initial magnetic field with respect to participant plane given by Monte Carlo Glauber simulation. (middle) AVFD predictions for CME-induced $H$-correlations in isobar collisions; (right) Predicted relative difference in $\gamma$-correlations in Zr-Zr and Ru-Ru collisions by folding together $F$- and $H$-correlations.} 
\label{fig.IsoBar}
\end{figure}

In Fig.\ref{fig.IsoBar} it shows the Monte-Carlo simulation result of the magnetic field, projected with respect to participant plane 
$B\equiv\langle B^2\cos(2\Psi_{\rm B}-2\Psi_{\rm PP}-\pi) \rangle^{1/2}$, in these systems, which differs for around $10\%$. 
Hence our AVFD simulation gives prediction that the CME signal would differs for $\sim 20\%$,  as shown in the middle panel.
Taking into account the non-CME background estimated from Au-Au collisions (details can be found in \cite{Shi:2017cpu}), we could still expect a relative difference around $10\%$. 
Given the sufficient statistic of the IsoBar program, such difference shall be observed.

\section{Summary}
\label{sum}
In summary, a new simulation tool --- the Anomalous-Viscous Fluid Dynamics (AVFD) framework has been developed for quantifying the charge separation signal induced by Chiral Magnetic Effect in relativistic heavy ion collisions. We find that,   
subject to current theoretical and experimental uncertainties, the AVFD-predicted CME signal with realistic initial conditions and magnetic field lifetime is quantitatively consistent with measurements from 200AGeV AuAu collisions at RHIC. Also, by event-by-event simulation, we find that the intercept of the $v_2$ dependence of the correlator $\gamma^{OS-SS}$ is directly   sensitive to the CME while the slope is likely to be dominated by background contributions.
Finally, we make predictions for the upcoming isobaric collisions that would be a critical test for the search of CME in heavy ion collisions.

 {\bf Acknowledgments.} 
This material is based upon work supported by the U.S. Department of Energy, Office of Science, Office of Nuclear Physics, within the framework of the Beam Energy Scan Theory (BEST) Topical Collaboration (YJ, JL, SS). The work is also supported in part by the NSF Grant No. PHY-1352368 (JL and SS), by the DFG Collaborative Research Center  ``SFB 1225 (ISOQUANT)'' (YJ), and by the NSF REU Program at Indiana University (EL).
The computation of this research was performed on IU's Big Red II \& Karst clusters,
that are supported in part by Lilly Endowment, Inc., through its support for the Indiana University Pervasive Technology Institute, and in part by the Indiana METACyt Initiative. The Indiana METACyt Initiative at IU was also supported in part by Lilly Endowment, Inc.

\end{document}